\documentclass[conference]{IEEEtran}

\usepackage[final]{graphicx}
\usepackage{graphicx} 
\usepackage{float}
\usepackage{lipsum}
\usepackage{pdfpages}
\usepackage{subcaption}
\usepackage{mathtools}
\usepackage[english]{babel}
\usepackage{amsthm}
\usepackage{thmbox}
\usepackage{bm}
\usepackage{mathrsfs}
\usepackage[scr=rsfs,cal=boondox]{mathalfa}
\usepackage{amssymb}
\usepackage{amsmath,tikz}
\usepackage{extarrows}
\usepackage{empheq}
\usepackage{adjustbox}
\usepackage[mode=buildnew]{standalone}

\usepackage[backend=biber,style=alphabetic,sorting=ynt ]{biblatex}
\def\miu{Miu}
\def\pname{FRAMER}

\newtheorem[M]{definition}{Definition}[section]
\newtheorem[M]{myequation}{Equation}[section]

\newcommand{\mm}[1]{$#1$}

\newcommand\slotmd[1]{s_{\mu}}

\usepackage{libertine}
\usepackage{epsfig}
\usepackage{tikz}
\usepackage{amsmath,tabu}
\usepackage{graphicx}
\usepackage{adjustbox}
\usepackage{amssymb,amsmath,enumitem}
\usepackage{fancyvrb}
\usepackage{color,soul}
\usepackage{array,colortbl,xcolor}
\usepackage{pgfplots}
\usepackage{pgfplotstable}
\usepackage{standalone}
\usepackage{url}
\usepackage{subcaption}
\usepackage{fancyvrb}
\usepackage{bussproofs}
\usepackage{setspace, xspace, multirow}
\usepackage{lipsum} 
\usepackage{xparse}
\usepackage{filecontents}

\usepackage{courier}
\usepackage[T1]{fontenc}
\usepackage[utf8]{inputenc}    
\renewcommand{\ttdefault}{cmtt}

\usepackage{xpatch}
\DeclareMathVersion{ttmath}
\SetSymbolFont{letters}{ttmath}{OT1}{\ttdefault}{m}{n}
\xapptocmd{\ttfamily}{\mathversion{ttmath}}{}{}

\usepackage{listings}
\def\BibTeX{{\rm B\kern-.05em{\sc i\kern-.025em b}\kern-.08em
    T\kern-.1667em\lower.7ex\hbox{E}\kern-.125emX}}

\newbox\verbbox
\usetikzlibrary{patterns}
\usetikzlibrary{decorations.pathmorphing, decorations.pathreplacing,calligraphy}
\usetikzlibrary{positioning,calc, fit}
\usetikzlibrary{shapes,arrows,shadows, shapes.multipart}

\definecolor{myorange}{RGB}{250,192,116}
\definecolor{rred}{HTML}{C0504D}
\definecolor{lgreen}{RGB}{204, 204, 154}
\definecolor{ppurple}{HTML}{9F4C7C}
\definecolor{ggreen}{HTML}{9BBB59}
\definecolor{bblue}{HTML}{4F81BD}
\definecolor{palepurple}{HTML}{F1DAFF}
\definecolor{palemint}{HTML}{CCFFDD}
\definecolor{palegreen}{HTML}{CCEEDD}
\definecolor{palered}{HTML}{FFCCCC}
\definecolor{pastelteal}{HTML}{A6DDCA}
\definecolor{pastelpurple}{HTML}{C7B7D9}
\definecolor{pink}{HTML}{FFF1F2}
\definecolor{skyblue}{HTML}{CCE5FF}

\tikzset{ 
    helpernode/.style={
        outer sep=0pt, 
        inner sep=0pt,
        node distance=0pt, 
        text width=1cm,
        minimum height=1cm,
        minimum width=1cm,
        trapezium stretches=true,
        rectangle, 
        text=white,
        draw,
    },helpernode/.default=1cm,
    docushape/.style={
        shape=tape,
        minimum width=5cm,
        inner ysep=3pt,
        draw,
        ultra thick,
        align=center,
        fill=white,
        font=\scriptsize\ttfamily, 
        tape bend top=none,
    },
    arrowstyle/.style={
        single arrow, 
        fill=blue!50, 
        anchor=base, 
        align=center,
        ultra thick,
        text width=2.8cm
    },
    mymodule/.style={
        rounded corners=0.5mm,
        draw=black,
        ultra thick,
        font=\footnotesize\ttfamily, 
        minimum width=2cm,
        minimum height=0.5cm,
    },
    myshade/.style={
       rounded corners=1mm,
       ultra thick,
       draw=gray,
       font=\scriptsize\ttfamily, 
       inner ysep=4pt,
       inner xsep=4pt
    },
    mypointer/.style={
       thick,
       draw=black,
       font=\scriptsize\ttfamily, 
    },
     box/.style={
        node distance=0pt,outer sep=0pt,
        draw=black,
        thick,
        fill=white, 
        minimum height=0.5cm,  
        font=\fontsize{8}{8}\ttfamily, 
        anchor=west,
    },
        object/.style={
        rounded corners=1mm,
        draw=black,
        fill=white, 
        minimum width=1.6cm,
        text height=2.0pt,
        text depth=0.5pt,
        font=\fontsize{6}{6}\ttfamily, 
        anchor=west
    },
    frame/.style={
        rounded corners=1mm,
        draw=black,
        thick,
        font=\scriptsize,
        minimum width=2cm,
        minimum height=0.3cm,
        inner sep=0pt
    },
    rightflat/.style={
        append after command={%
            \pgfextra
            \fill[fill=#1] 
            (\tikzlastnode.south east) 
            [rounded corners] -| 
            (\tikzlastnode.west) |- 
            (\tikzlastnode.north) 
            [sharp corners] -| cycle;
        \draw[rounded corners] 
            (\tikzlastnode.south east) -| 
            (\tikzlastnode.west) |- 
            (\tikzlastnode.north east);
            \endpgfextra},
        minimum width=2cm,
        inner sep=1.8pt,
        font=\scriptsize,
        thick
    },
    leftflat/.style={
        append after command={%
            \pgfextra
            \fill[fill=#1] 
            (\tikzlastnode.north west) 
            [rounded corners] -| 
            (\tikzlastnode.east) 
            |- (\tikzlastnode.south) 
            [sharp corners] -| cycle;
        \draw[rounded corners] 
            (\tikzlastnode.north west) -| 
            (\tikzlastnode.east) |- 
            (\tikzlastnode.south west);
        \endpgfextra},
        minimum width=2cm,
        text height=2.0pt,
        text depth=0.5pt,
        fill=bblue,
        thick
    },
    entry/.style={
        node distance=0pt,outer sep=0pt,
        draw=black,
        text width=#1,
        font={\sffamily\bfseries},
        align=center,
        minimum width=0.1cm,
        minimum height=0.3cm
    },
     bits/.style={
        draw=black,
        text width=#1,
        font={\sffamily\bfseries},
        align=center,
        text height=2.4pt,
        text depth=0pt,
        font=\fontsize{6}{6}\ttfamily, 
        anchor=west
    }
}

\pgfplotsset{
  bargraph/.style={
    xlabel={},
    ylabel={},
    xlabel style={font=\tiny},
    ylabel style={align=center, text width=2.2cm, 
                  font=\sffamily\tiny},
    axis x line*=bottom,
    axis y line*=right,
    enlarge x limits=0.05,
    set layers,
    ybar=0.0pt,
    bar width=0.08cm,
    width=6cm, height=2.3cm, 
    ymin=0, 
    symbolic x coords={
        perlbench, bzip2, mcf, hmmer, sjeng, 
        libquantum, h264ref, omnetpp, astar, xalancbmk	 
    },
    xtick={
        perlbench, bzip2, mcf, hmmer, sjeng, 
        libquantum, h264ref, omnetpp, astar, xalancbmk	 
    },
    visualization depends on=rawy\as\rawy,
    clip=false,
    legend cell align=left,
    x label style={font=\tiny},
    y label style={font=\tiny},
    xticklabel style={font=\fontsize{4}{4}\bfseries,
        rotate=30,
        anchor=east},
    yticklabel style={font=\fontsize{5}{5}\bfseries},
    ylabel near ticks,
    ymajorgrids, major grid style={draw=gray},
    legend style={
        draw=none,
        fill=none,
        at={(0.85, 0.75)},
        anchor=south, legend columns=-1, 
        font=\fontsize{4.5}{4.5}\ttfamily
    },
    xtick align=inside
  }
}

\def\myheight{3.0cm}
\pgfplotsset{
  bargraphcsv/.style={
    xlabel={},
    ylabel={},
    table/col sep=comma,
    axis x line*=bottom,
    axis y line*=right,
    enlarge x limits=0.05,
    set layers,
    ybar=0.0pt,
    bar width=0.08cm,
    width=6cm, height=\myheight, 
    ymin=0, 
    visualization depends on=rawy\as\rawy,
    clip=false,
    legend cell align=left,
    x label style={font=\tiny},
    y label style={font=\tiny},
    xticklabel style={font=\fontsize{4}{4}\bfseries,
        rotate=30,
        anchor=east},
    yticklabel style={font=\fontsize{5}{5}\bfseries},
    ylabel near ticks,
    ylabel style={align=center, text width=\myheight, 
                  font=\sffamily\tiny},
    ymajorgrids, major grid style={draw=gray},
    legend style={
        draw=none,
        fill=none,
        at={(0.8, 0.90)},
        anchor=south,
        legend columns=-1,
        font=\fontsize{4.5}{4.5}\ttfamily
    },
    xtick align=inside
  }
}

\tikzstyle{arrow}=[draw, -latex]

\def\bitwidth_dis{0.5cm}

\pgfplotsset{compat=newest}

\pgfkeys{
    /tikz/node distance/.append code={
        \pgfkeyssetvalue{/tikz/node distance value}{#1}
    }
}

\def\miu{Miu}
\def\pname{FRAMER}

\newcommand{\mycomment}[1]{}

\begin{document}

\title{\textbf{FRAMER/Miu}: Tagged Pointer-based Capability \\and \\ Fundamental Cost of Memory Safety \& Coherence \\
(Position Paper)}




%
\author{
    \IEEEauthorblockN{Myoung Jin Nam \IEEEauthorrefmark{1}}
    \IEEEauthorblockA{ University of the West England \\
        Email: myoung-jin.nam@uwe.ac.uk, mjn31@cantab.ac.uk}
}



\maketitle

\begin{abstract}

Ensuring system correctness, such as memory safety, can eliminate security vulnerabilities that attackers could exploit in the first place. However, high and unpredictable performance degradation remains a primary challenge.

Recognizing that it is extremely difficult to achieve complete system correctness for production deployment, researchers make trade-offs between performance, detection coverage, interoperability, precision, and detection timing.

This research strikes a balance between comprehensive system protection and the costs required to obtain it, identifies the desirable roles of software and hardware, and presents a tagged pointer-based capability system as a stand-alone software solution and a prototype for future hardware design. This paper presents follow-up plans for the \pname{}/\miu{} generic framework to achieve these goals.


\end{abstract}


%
\IEEEpeerreviewmaketitle




\section{Introduction}\label{sec:introduction}

\hfill 27th August, 2024





One defense mechanism is to allow memory errors to occur but harden the program to prevent exploitation by implementing techniques such as Control-flow Integrity (CFI)~\cite{cfi, llvm_cfi_3.9, Zhang2016VTrustRT,ccfir,bincfi} and Address Space Layout Randomization (ASLR)~\cite{aslr}. Another mechanism is to ensure \textit{system correctness} by detecting and blocking security vulnerabilities within a system from the outset. For example, memory safety enforcement can prevent 70-80\% of attacks that exploit memory errors~\cite{chrome.vulnerability.memory.error}.

When a program accesses data in memory during execution, the only information available is the memory address that a pointer holds. How do we check if the memory access is valid—whether the address is within a valid range, the intended object has not been freed, the interpretation of data is correct, the pointer has the right to access the data, or the access is in the correct execution order?

One of the inevitable and dominant approaches is to check individual accesses during execution. Unfortunately, run-time checking is extremely resource-intensive, mainly due to the need to update and track run-time metadata—where to store metadata and how to locate it with an object pointer. Metadata retrieval at run-time is one of the major contributors to performance degradation, making efficient metadata management a key challenge.


This research begins with the premise that it is nearly impossible to ensure comprehensive and complete system correctness at a performance level suitable for production deployment (overhead < 5\%) solely with software solutions. Hardware acceleration is necessary for an \textit{always-on} solution. To this end, \miu{} strikes a balance between performance, interoperability, detection coverage, precision, and detection timing. The performance matrix can be broken down into two major assessment criteria—run-time and memory overheads. We push the limits of performance metrics to the extreme, depending on the deployment and goal, and enforce system correctness and security.

\begin{figure}[tbp!]
    \centering
        \includegraphics[width=\linewidth, trim=8.5cm 6cm 9cm 6cm, clip]{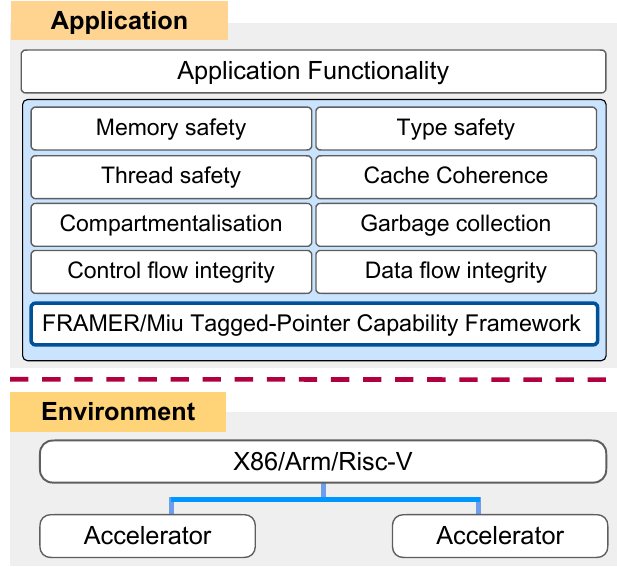}
    \caption{FRAMER/Miu framework}
    \label{fig:miu.architecture}
\end{figure}



In this paper, we focus on \textit{memory safety} and \textit{coherence}. By \textit{safe}, we mean that access rights for individual loads/stores are authorized in association with objects and pointers at runtime. By \textit{coherent}, we mean that the execution of a system with non-coherent hardware accelerators remains correct. Among the two major approaches within the \miu{} framework that use \textit{sideline} (concurrent) monitoring~\cite{MemPatrol}, this paper primarily reviews \textit{inline} reference monitoring, \pname{}/\miu{}, and discusses the challenges and plans for extending it to ensure coherence~\cite{guy.cache.abstraction}.

The basic idea of \pname{}~\cite{FRAMER_ARM, FRAMER, jin.thesis.technical.report} is to place \textit{per-object metadata} in a location whose address can be derived solely from a \textit{tagged pointer}. \pname{} attaches metadata to an object as a header and calculates the lower bound from the derived metadata location for greater cache efficiency. \pname{} locates and retrieves metadata during a run-time check to authorize memory access based on the result. A tagged pointer holds the \textit{relative location} of metadata along with an arbitrary address. The address space does not change, and the tag size is adjustable. Note that the destination of derivation is not necessarily the metadata location; it can be any location that needs to be calculated from a tagged pointer, such as the lower or upper bounds of an object.

\begin{figure}[tbp!]
    \centering
    \includegraphics[width=\linewidth, trim=5.3cm 3.1cm 4.9cm 2.8cm, clip]{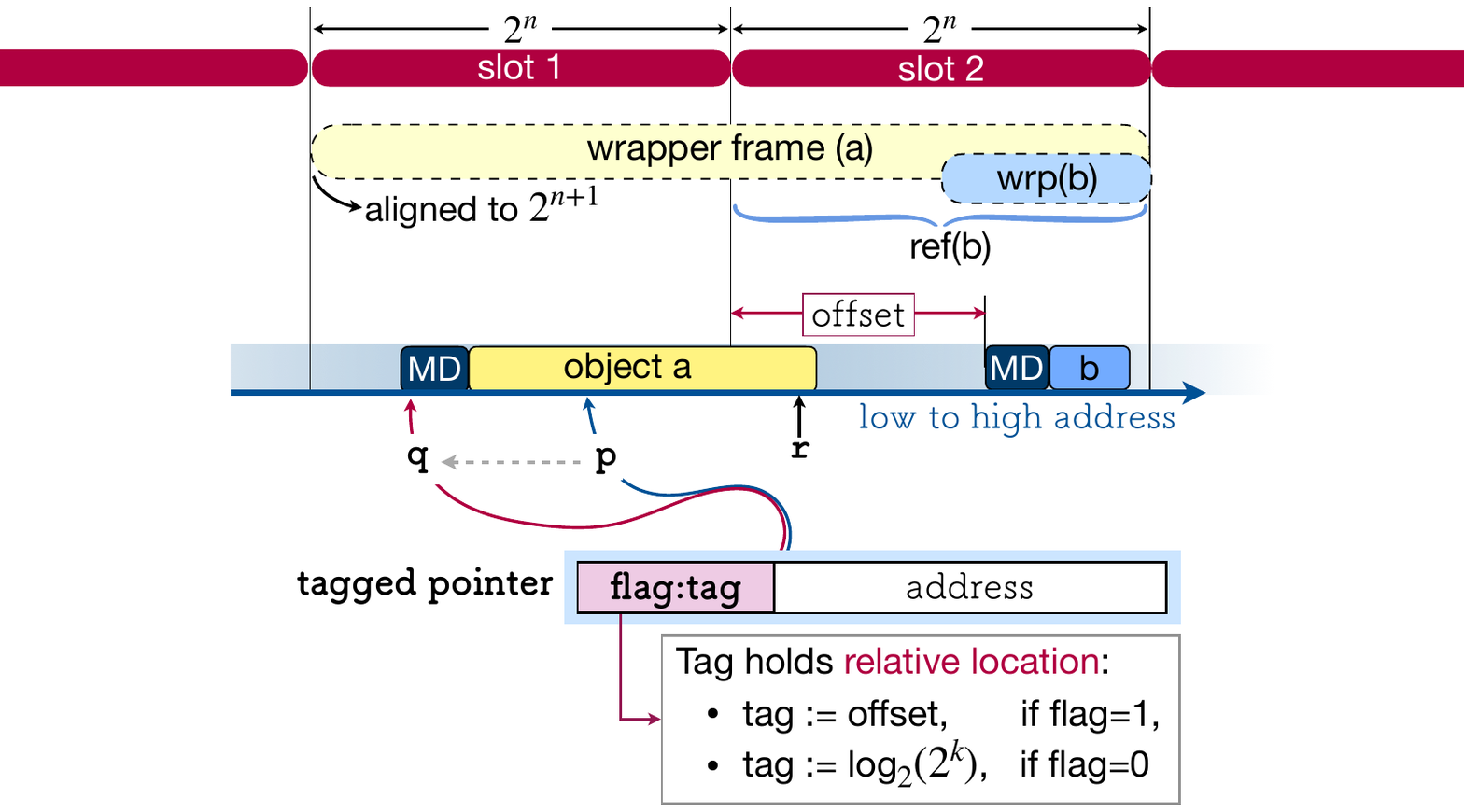}
    \caption{FRAMER's wrapper frame and tagged pointer. FRAMER places per-object metadata (\textbf{MD}) near its object and tags a pointer with (1) a flag indicating the object's wrapper frame size (\texttt{ON}, if smaller than the slot size) and (2) a tag holding the relative location to \textbf{MD}. During memory access, FRAMER derives a metadata pointer (\textbf{q}) solely from a tagged pointer \textbf{p}, which contains an arbitrary address and tag.
}
    \label{fig:wrapper.frame.diagram}
\end{figure}

\pname{}/\miu{} is designed with three major focuses:

\begin{itemize}

\item \textbf{High Performance:} \miu{} aims to minimize the \textit{fundamental costs}—the overhead unresolved by hardware support, such as slowdowns caused by cache misses or memory overhead. The slowdown in \pname{}'s software implementation resulting from an increase in executed arithmetic operations can be largely resolved with a single customized instruction.

\item \textbf{Architecture Independence:} Supported by its innovative pointer encoding, this framework is adaptable across various architectures, enabling immediate adoption at a low cost. \miu{} found a solution for full compatibility through ARMv8's system-wide support of Top Byte Ignore (TBI). A similar resolution is expected for x86, considering both AMD and Intel have adopted tagged pointers. 

\item \textbf{Scalability:} The system scales to various policies encompassing memory safety, type safety, compartmentalization, garbage collection, thread safety, and cache coherence. Currently, it provides spatial memory safety and type confusion checking, which are the main targets of attackers for exploitation.

\end{itemize}




While providing fine-grained and comprehensive protection, as shown in Fig.~\ref{fig:miu.architecture}, \pname{}/\miu{} aims to minimize the fundamental costs. Many memory safety solutions tolerate an increase in memory usage for better speed, since memory safety enforcement has been used mainly for troubleshooting during development, where memory resources are not the primary concern. However, this approach invites debate for production deployment. There are systems where memory resources are as critical as speed, such as embedded systems with limited memory space or I/O server systems. In addition, some contributors to slowdown can be easily resolved with hardware acceleration, e.g., customized instruction sets, while memory overhead cannot be resolved with hardware support.

This paper will overview the prior \pname{} mechanism in Section~\ref{sec:overview.framer}, and then discuss challenges and plans in  Section~\ref{sec:low.cost.memory.safety.coherence} and Section~\ref{sec:extension.of.security.policies}. Section~\ref{sec:related.work.memory.safety} describes general preliminaries of memory safety and related work on unconventional pointer representations and \mm{2^n}-based tag encoding.



\section{Overview: FRAMER}
\label{sec:overview.framer}




\begin{figure}[]
    \centering
    \includegraphics[width=\linewidth, trim=7.2cm 9.5cm 7.2cm 3cm, clip]{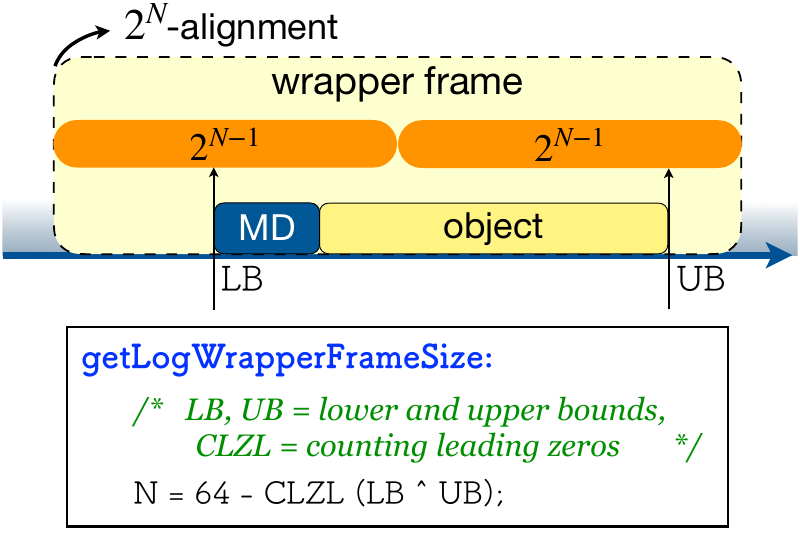}
    \caption{Calculation of a wrapper frame for an object with lower and upper bounds}
    \label{fig:cal.wrapper.frame}
\end{figure}

The key contribution of \pname{} is the improvement of prior \mm{2^n}-based encoding of relative location~\cite{baggybounds} by enabling natural alignment, i.e., eliminating the need for object re-alignment to a larger \mm{2^n} and avoiding object padding. As shown in Fig.~\ref{fig:wrapper.frame.diagram}, a per-object metadata pointer (\textbf{q}) to an object \textbf{a} is derived solely from a tagged pointer (\textbf{p}), which contains an arbitrary address and a relative location. Each metadata (\textbf{MD}) is placed near its object for cache efficiency, without altering the internal memory layout of the objects. Notably, \pname{} has demonstrated:

\begin{itemize} 
    \item Low memory overhead (23\% vs. 784\%) compared to AddressSanitizer (ASan)\cite{address-sanitizer}. 
    \item Superior cache efficiency (cache misses: 40\% vs. 100-130\%) compared to ASan and Cheri\cite{cheri.TR.2023}. 
    \item Near-zero false negatives, surpassing ASan, Cheri, and Arm’s MTE~\cite{arm85a}, all of which exhibit false negatives. 
\end{itemize}

\pname{} introduces a \textit{wrapper frame} concept to encode relative locations. \pname{} first defines a \textit{frame} as a \mm{2^n}-sized memory block which is aligned to its size. 
Every memory object $O$ is completely contained (\textit{bounded}) \textit{at least} within one frame. Among $O$'s bounding frames, $O$'s \textit{wrapper frame} is defined as the smallest bounding frame of $O$. For instance, Fig.~\ref{fig:wrapper.frame.diagram} shows objects \textbf{a} and \textbf{b} whose wrapper frame sizes are \mm{2^{n+1}} and \mm{2^{n-1}}, respectively. Since $w_O$ is the smallest bounding frame, there is only one wrapper frame for each object, and it does not change in location or size during the lifetime of \textit{O}. \pname{} encodes relative location to metadata using this wrapper frame information. Fig.~\ref{fig:cal.wrapper.frame} illustrates how to calculate a wrapper frame size from lower and upper bounds, which are known at memory allocation.

\subsection{Tag encoding and decoding}
\label{sec:tag_encoding_at_alloc}

Unfortunately, the top spare bit space is limited and cannot accommodate both wrapper frame information and metadata location. To address this, \pname{} logically divides the virtual address space into fixed-sized \mm{2^n}-frames, called \textit{slots}, where \mm{n+1} is the spare bit width. The last spare bit is reserved for the \texttt{flag}. This approach allows \pname{} to allocate some of the tag space for wrapper frame information.

\pname{} categorizes objects into two types based on the wrapper frame size. If a metadata-attached object is contained within a slot, i.e., \textit{small-framed}, \pname{} tags only the offset of metadata within the slot (also referred to as a reference frame of an object) with \texttt{flag=1}. During run-time checks, metadata can be located by zeroing out the least significant \textit{n} bits of a pointer and then adding the offset. In Fig.~\ref{fig:wrapper.frame.diagram}, a pointer to a small-framed object \textbf{b} is tagged with an offset from \textbf{slot 2} to \textbf{b}'s \textbf{MD}, with \textbf{flag=1}.

If an object's wrapper frame is larger than a slot, this calculation does not work because pointers can legally go beyond the slot boundaries, e.g., a pointer \textbf{r} to a large-framed object \textbf{a} in Fig.\ref{fig:wrapper.frame.diagram}. Such pointers would be incorrectly derived if only the offset is used. In Fig.\ref{fig:wrapper.frame.diagram}, a pointer \textbf{r} to object \textbf{a} is incorrectly derived as being within \textbf{slot 2}. Additionally, using an offset is impractical for large wrapper frames because it does not fit within the tag space. For large-frame objects, \pname{} stores the metadata address in a shadow table and tags the object pointer with \mm{N=\log_2}(\text{wrapper frame size}). During memory access, \pname{} locates the corresponding entry holding the metadata address using the \mm{N} value. For example, pointers to an object \textbf{a} are tagged with \textit{N+1}.

\begin{figure}[]
    \centering
    \includegraphics[width=\linewidth, trim=5.3cm 4.5cm 5cm 3.8cm, clip]{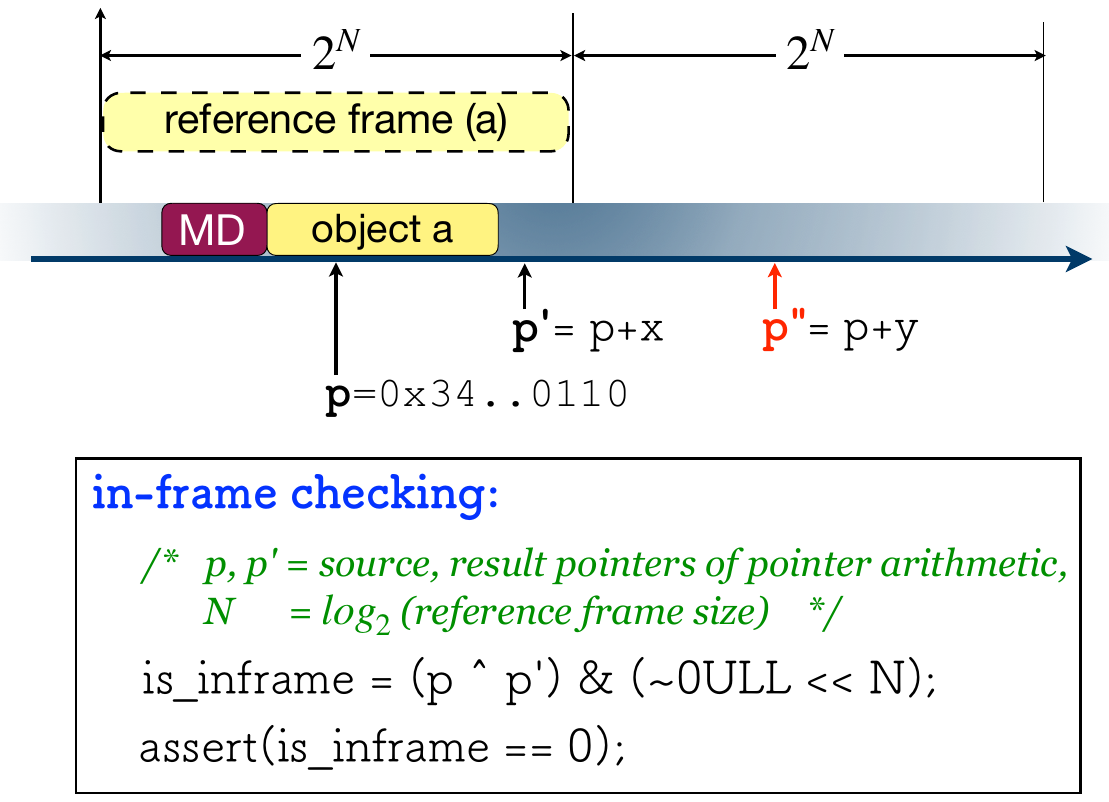}
    \caption{FRAMER's in-frame checking at pointer arithmetic. To detect pointers going out of reference frame (e.g., a pointer \textbf{p"}), FRAMER performs XOR operation with a source and result pointer at pointer arithmetic.}
    \label{fig:framer.inframe.checking}
\end{figure}

\label{sec:tag_encoding_at_access}



\subsection{In-frame Checking at Pointer Arithmetic}
\label{sec:inframe_checking_at_pointer_arithmetic}


\pname{} uses per-object metadata and relative location encoding, requiring to address the issue of pointers going out-of-bounds of their \textit{intended referents}~\cite{jones97backwards}. \pname{} performs \textit{in-frame checking} to determine whether a pointer goes out of its reference frame (such as a reference slot or wrapper frame) during pointer arithmetic , as illustrated in Fig.\ref{fig:framer.inframe.checking}. The advantage of this technique is its simplicity: it relies on a straightforward arithmetic operation (XOR) and does not require access to metadata.

\section{Challenge: Low-cost Memory Safety and Coherence}
\label{sec:low.cost.memory.safety.coherence}

\subsection{New Tagged Pointer Compression}



Before considering performance, it is essential to ensure compatibility within the environment. \pname{} ensures that store/load operations are executed with a tag-cleaned pointer after runtime checks to prevent segmentation faults during pointer dereferences on X86. Additionally, \pname{} guarantees interoperability between instrumented modules and precompiled libraries by using regular tag-free pointers, stripping tagged pointers before passing them to non-instrumented functions on X86. Unfortunately, there is a rare case where \pname{} cannot clear the tag through argument instrumentation -- tagged pointers may escape from instrumented functions and potentially cause segmentation faults in uninstrumented code.

The runtime overhead for ensuring compatibility through software alone is not negligible (10-15\%). Moreover, unless individual memory accesses are proven safe at compile time, over-instrumentation of memory access may be necessary to avoid segmentation faults.

\subsubsection{Leveraging Top Byte Ignore (TBI) in Armv8}

In this context, Armv8's Top Byte Ignore (TBI) feature becomes very attractive~\cite{arm85a}. TBI masks the top 8 bits in a pointer during memory access, thus supporting system-wide tagged pointers. This approach eliminates the need for software-based tag-cleaning. We modify \pname{}/\miu{} to leverage TBI in Armv8 as an initial step.

To utilize TBI, the tag size, including the flag bit, must fit within 8 bits. The first challenge is that as the tag space becomes smaller, the shadow table for large-framed objects grows larger. The slot size is determined by the number of tag bits, with each slot mapped to one table entry (e.g., one entry per \mm{2^{15}} with a 15-bit relative location). Therefore, reducing the slot size increases the number of entries, leading to greater memory consumption.

Another challenge is that a smaller slot size results in a higher number of \textit{small-sized} ($\leq$ slot size) but \textit{large-framed} objects. While such objects are uncommon with 16-bit spare bits, a significantly smaller tag size (e.g., 8 bits including a flag) categorizes many small-sized but large-framed objects. These objects require expensive accesses to entries in a remote region because the metadata of large-framed objects cannot be located solely from tagged pointers. This situation significantly increases both runtime and memory overhead.

Additionally, the tag space in all architectures is decreasing; for example, Intel utilizes address spaces larger than \mm{2^{48}}. To address this, \miu{} requires new encoding strategies for immediate adoption in new architectures, including Armv8 with the TBI feature, to maintain low overhead.

\subsection{Customised Instruction Set Extensions}

\pname{} mechanism is at its best when it is implemented as instruction set extensions (ISA). 



The focus is on eliminating overhead that hardware support cannot resolve, aiming for higher efficiency in data cache usage and memory footprint. To achieve this, \pname{} compromises on optimized dynamic instruction counts. Specifically, \pname{} (1) minimizes the increase in extra cache misses for metadata (due to spatial locality) and (2) tolerates an increase in executed instructions for arithmetic operations. The primary contributor to the slowdown in \pname{} is the sequence of arithmetic operations required during memory access. Generating a tag and deriving a metadata address can each be implemented as a single operation.

\pname{} has demonstrated better performance than might be expected. The evaluation shows excellent D-cache performance, with the impact of software checking being, to a significant extent, mitigated by improved instructions per cycle (IPC).

\section{Challenge: Extension of Security Policies}
\label{sec:extension.of.security.policies}


\miu{} is a generic framework designed to accommodate various security policies within a unified capability system, as shown in Fig.~\ref{fig:miu.architecture}. \miu{}'s per-object metadata and table entries can hold any information and can be utilized for other policies as well. Currently, it provides spatial memory safety and type confusion checking, which are primary targets for attackers. The framework's scope will gradually expand to include garbage collection, detection of dangling pointers, \textit{happens-before} relation checking in shared memory and cache coherence in heterogeneous systems with non-coherent accelerators. This will require finer-grained and temporal information, such as states. We briefly discuss two of these policies in this section.


\subsection{Type Confusion Checking in C/C++}



Type conversion in C/C++~\cite{ccured,typesan} is often necessary despite its safety risks. We ensure that type conversions make sense, i.e., they do not cause data loss, incorrect interpretation of bit patterns, or memory corruption. Type confusion is frequently combined with dangling pointers in attacks, where the memory area of a de-allocated object (the old object) is reused by a new object. The type mismatch between the old and new objects can allow an attacker to access unintended memory.

Run-time type confusion is a direct application domain of \pname{}/\miu{}. One challenge in run-time typecast checking is \textit{pointer-to-type} mapping. When pointers are typecast to different types, they may be offset to one of the sub-fields of the object. Therefore, the pointer must be mapped to the corresponding type at that offset. This requires associating each object (or pointer) with its object type and mapping a pointer to the object’s type using the type information stored in metadata. Unfortunately, this process incurs high overhead. The pointer must then be examined to ensure that the type conversion from the type at the offset to the target type is safe.


Type checking on x86 has been published  ~\cite{jin.thesis.technical.report}. The next step is to update the tagged pointer representation and metadata management, as suggested in Section~\ref{sec:low.cost.memory.safety.coherence}, for ARMv8 with the TBI feature.

\subsection{Coherence Checking in Multi-threaded Systems}



Hardware accelerators are typically \textit{non-coherent} to achieve simpler design and higher performance. Traditional hardware-based and full software-based coherence methods are overly strict, enforcing coherent actions on all accesses. Even with small workloads, data flushing can consume over 50\% of run-time with non-coherent accelerators, making them up to three times slower~\cite{accelerators.and.coherence.soc.perspective}.
Therefore, it is desirable to avoid excessive flushing and reduce run-time checks to the minimal subset. By analyzing access patterns, we can instrument only what is necessary to maintain performance and correctness.

\miu{} can be developed to integrate both software and hardware methods to ensure both memory safety and coherence, as both policies require tracking load and store addresses to detect memory safety violations or enforce coherence actions~\cite{guy.cache.abstraction}. To keep run-time overhead low, both software methods (such as formal static analysis) and hardware methods (such as acceleration) must be suitable for production code. The software methods involve a combination of formal static analysis and code generation to instrument only the necessary loads and stores, while hardware methods accelerate these instrumented checks.


\section{Related Work: Memory Safety }
\label{sec:related.work.memory.safety}


This section reviews prior approaches to metadata management for spatial memory safety and discusses the trade-offs involved in the dynamics of detection coverage, precision, and performance. Our focus is on enforcing memory safety through low-level, \textit{inline reference monitors} that embed checks to prevent memory errors by instrumenting existing unsafe code.

\subsection{Metadata management}
\label{sec:metadata.management}


Memory safety solutions can be divided into two categories depending on whether they associate metadata (such as bounds information) with each pointer or memory object.

\subsubsection{Pointer-based Tracking}
\label{sec:pointer_tracking}

Approaches in this category associate each \emph{pointer}
with metadata i.e. address range that the pointer is allowed to point to~\cite{nagarakatte_et_al:LIPIcs:2015:5026}.
Therefore, per-pointer metadata can hold bounds at \emph{byte-granularity}, permitting creation of \emph{out-of-bounds} pointers and pointers to sub-objects that are allowed in C/C++. This makes it easier to detect \emph{internal overflows} such as an array out-of-bounds inside a structure. 

The most straightforward and conventional per-pointer metadata is \textit{fat pointers}~\cite{Austin:1994:EfficientDetection,
Jim_cyclone:a, ccured, Chisnall:2015:BPA:2694344.2694367}. 
This expended pointer representation carry metadata with itself. Conventional fat pointers~\cite{ccured, old_cheri, Jim_cyclone:a} store a range of \textit{absolute} addresses in full bit-width to per-pointer metadata at pointer assignment and perform bounds checking \emph{only} at memory access without extra operations at pointer arithmetic. This allows fine-grained bounds checking with near zero false negative/positive. However they lose benefits of per-pointer metadata, when the approaches compress bounds information such as offset-based \textit{relative} location or manipulating memory alignment~\cite{cheri.TR.2023}  (discussed later in Sec~\ref{sec:object_tracking}).

Due to the increased pointer size, fat pointers unfortunately require memory layout modification, damaging interoperability with non-instrumented code. Therefore per-pointer metadata are often assisted with hardware~\cite{arm.morello, cheri_risc, intel_mpx}, while software-only solutions dominantly implement per-object metadata.  Another disadvantage is D-cache impact of increased pointer size, resulting in non-negligible performance degradation~\cite{capvm}.

Some approaches ~\cite{MSCC, softbound, 
intel_mpx, nagarakatte_et_al:LIPIcs:2015:5026} 
decouple metadata from a pointer representation and manage \textit{disjoint metadata} to ensure binary compatibility as trade-off of higher locality than fat pointers. Unfortunately, disjoint per-pointer metadata incur significant overhead. Especially the number of pointers is typically larger than that of allocated objects, so pointer-intensive programs may suffer from heavier runtime and memory overheads. 

Pointer-tracking approaches have additional runtime overhead from metadata copy and update at pointer assignment, while object-tracking approaches create metadata only at memory allocation/release. 

It is still difficult to achieve full compatibility
with them -- if a pointer created within an instrumented module is passed to and modified in an un-instrumented module, the corresponding metadata is not updated, causing false violations.

\subsubsection{Object-based Tracking}
\label{sec:object_tracking}

Many software-based techniques ~\cite{baggybounds,
address-sanitizer,
Dhurjati:2006:BAB:1134285.1134309,
jones97backwards,
effectivesan,
PAriCheck}
store metadata \emph{per object}. 
They provides great compatibility with current source code and pre-compiled legacy libraries by not changing the memory layout of objects. In addition, per-object metadata is updated only at memory allocation/release so even if a pointer is updated in an un-instrumented module, the metadata does not go out-of-sync. 

Per-object metadata checks bound at object granularity, so it requires additional care for finer-grained checking e.g.  internal overflows, while pointer-tracking approaches can simply assign an address range of a sub-object to a pointer.

Early object-tracking approaches~\cite{jones97backwards} store an address range of each object in a data structure, so they require expensive \emph{address range lookups}, while pointer-tracking allows access the corresponding entry using the address of a pointer as a \emph{key} in the metadata table. 

To address this, software-based solutions dominantly adopted \textit{shadow space}~\cite{baggybounds,PAriCheck,FRAMER,
address-sanitizer, metalloc}. 
Basic idea of Baggy Bounds (BBC)~\cite{baggybounds} and Address Sanitizer (Asan)~\cite{address-sanitizer} is to logically divide virtual address space into \textit{slots} (fixed \mm{2^n}-sized blocks aligned to the size); mandates object alignment to \mm{2^n}; and store \textit{per-slot} metadata in one byte-sized entries in the shadow space. \mm{2^n}-alignment of objects aims to prevent metadata conflicts caused by multiple objects in one slot. Although inflated object size significantly increases memory overhead, shadow space has great advantage -- single direct array access to metadata by calculation using mapping ratio, removing expensive traverse in a data structure. With its usefulness in information compression,  \mm{2^n}-based encoding has been adopted by recent unconventional pointers, described in Sec~\ref{sec:prelimiary.pow2.encoding}.  

Another challenge of per-object metadata is handling pointers going out-of-bounds of a \emph{right} object (\emph{intended referent})~\cite{jones97backwards} by pointer arithmetic and landing in valid range of \emph{another} object. Memory access with these pointer can be seen valid, as long as the pointers are in-bounds of \textit{any} objects.  Therefore object-tracking approaches should take a special care to keep track of intended referents. 
The early approach J\&K~\cite{jones97backwards} pads objects with extra one byte (\emph{off-by-one byte}), and ASan adds extra pads (redzone) between objects, in addition to padding for \mm{2^n}-alignment, increasing memory overhead. Padding still can cause false results, when a pointer legitimately goes beyond padded area. 
BBC~\cite{baggybounds} instead performs bounds checking only at pointer arithmetic and marks the out-of-bounds pointers, so that errors are detected at pointer dereferences. \pname{} performs one bit operation at pointer arithmetic to just check if a pointer stays in a derivable region-- removing a memory waste of padding and extra metadata retrieval. Both approaches embed false positives where an illegal pointer comes back in-bounds without being dereferenced. 


\subsection{Unconventional Pointers and Relative Location Encoding}
\label{sec:prelimiary.pow2.encoding}



Widely-used compact shadow space schemes~\cite{baggybounds, address-sanitizer} utilize metadata for each \mm{2^n}-sized memory block\footnote{In general, every object is re-aligned to a fixed \mm{2^n} size greater than the maximum alignment.}, rather than for each object (or pointer). Each entry in the shadow space holds encoded metadata for the corresponding block. Baggy Bounds Checking~\cite{baggybounds} stores \mm{log_{2}}(padded object size) in a one-byte-sized entry, which enables bounds checking at the \textit{padded\_object} granularity by tolerating pointers going out-of-bounds but still within the padded bounds. ASan~\cite{address-sanitizer} improves on BBC with a different encoding in the entries, allowing for  object-grained precise bounds checking.


Unconventional pointers adopt \mm{2^n}-based encoding~\cite{arm85a, cheri.TR.2023, in.fat.pointer, low-fat-pointer}. Low-fat pointers~\cite{low-fat-pointer} extend Baggy Bounds' design for pointer compression and bounds checking, and Cheri~\cite{cheri.concentrate.compression, cheri_hybrid} compresses 256-bit fat pointers into 128 bits based on Low-fat pointers. Due to the limited number of spare bits in a pointer, 128-bit Cheri and Low-fat pointers re-align large objects and perform \textit{approximate} checking at the \textit{padded\_object} granularity. Additionally, 128-bit Cheri loses the advantage of per-pointer metadata holding the absolute location. \mm{2^n}-based relative location requires extra checks during pointer arithmetic, similar to object-tracking approaches, to keep track of (1) intended referents or (2) derivable address space with relative location. This additional challenge may be inevitable to reduce fat pointer size or utilize space in a pointer.



MTE~\cite{arm85a} uses a coloring scheme~\cite{WIT} and mandates smaller alignment (\mm{2^4}). MTE is similar to shadow space schemes, holding per-slot metadata but with hardware support (tagged memory). It performs probabilistic checking with a 1/16 chance of a false negative at every memory access.

In-fat pointers~\cite{in.fat.pointer} are hardware-assisted tagged pointers that provide byte-grained bounds checking on objects. They leverage object type information and update tag bits during pointer arithmetic.
Delta pointers~\cite{deltapointers} use an offset-based encoding, not \mm{2^n}-based. They tag a pointer with an offset to the upper bound, and update the offset during pointer arithmetic. Due to the limited space for holding offsets, Delta pointers provide exact checking only for small-sized objects.


Tagged pointers that update tag information during pointer arithmetic may occasionally produce false results if pointers are converted to integer types and undergo arithmetic operations before being dereferenced. \pname{} aims to avoid updating tag encoding at pointer arithmetic to minimize false negatives and false positives. However, we remain open to updating pointer arithmetic in future designs if it proves beneficial.

Unconventional pointers often face challenges in avoiding extra memory space beyond what is allocated in the pointer itself. Tagged pointers use less space but may require compromises in detection coverage, such as limiting detection to specific types of errors or employing probabilistic checking. For example, 128-bit Cheri pointers currently provide padded bounds checking for large objects but can incur additional memory overhead due to increased pointer size and greater alignment requirements (\mm{2^4}) for both pointers and large objects. These requirements can significantly impact performance, leading to an increase in D-cache misses by up to 90\%–100\% compared to Cheri’s 64-bit pointers in hybrid mode. \pname{} maintains a constant pointer size and manages per-object metadata to support security policies beyond mere bounds checking.

\label{sec:challenges}

\section{Conclusion}


\pname{}~\cite{FRAMER}/\miu{} is a hybrid model that integrates tagged pointers with disjoint metadata to ensure precise bounds checking for objects of all sizes and to provide scalable security enforcement. Our goal is to extend \miu{} to support a variety of security policies within this framework while refining the design to minimize the fundamental costs associated with system correctness. By addressing vulnerabilities proactively, \miu{} aims to ensure both security and reliability from the outset.


\printbibliography

\end{document}